\documentclass[fleqn,usenatbib]{mnras}

\usepackage{newtxtext,newtxmath}
\usepackage[T1]{fontenc}

\DeclareRobustCommand{\VAN}[3]{#2}
\let\VANthebibliography\thebibliography
\def\thebibliography{\DeclareRobustCommand{\VAN}[3]{##3}\VANthebibliography}

\usepackage{graphicx}	% Including figure files
\usepackage{amsmath}	% Advanced maths commands
\usepackage{comment}
\usepackage{float}
\newcommand{\msun}{\mathrm{M}_\odot}
\newcommand{\mstar}{M_\star}
\newcommand{\subfind}{\textsc{Subfind}}

\newcommand{\modtextv}{\color{black}}
\newcommand{\mstarpeak}{M_{\star,\,\mathrm{peak}}}
\newcommand{\vmaxpeak}{v_{\mathrm{max, peak}}}

\title[Early self-quenching of cluster galaxies at $z \geq 1.5$]{An environment-dependent halo mass function as a driver for the early quenching of $z\geq1.5$ cluster galaxies}
\author[S. L. Ahad et al.]{Syeda Lammim Ahad,$^{1, 2}$\thanks{E-mail: ahad@strw.leidenuniv.nl}
Adam Muzzin,$^{3, 1}$ Yannick M. Bah\'{e},$^{4, 1}$ and
Henk Hoekstra$^{1}$
\\
$^{1}$Leiden Observatory, Leiden University, P.O. Box 9513, 2300 RA Leiden, The Netherlands\\
$^{2}$Waterloo Centre for Astrophysics, University of Waterloo, Waterloo, ON N2L 3G1, Canada\\
$^{3}$Department of Physics and Astronomy, York University, 4700, Keele Street, Toronto, ON, MJ3 1P3, Canada\\
$^{4}$Institute of Physics, Laboratory of Astrophysics, Ecole Polytechnique F\'{e}d\'{e}rale de Lausanne (EPFL), Observatoire de Sauverny, 1290 Versoix, Switzerland
}

\date{Accepted XXX. Received YYY; in original form ZZZ}

\pubyear{2023}

\begin{document}
\label{firstpage}
\pagerange{\pageref{firstpage}--\pageref{lastpage}}
\maketitle

\begin{abstract}
Many $z$$\approx$1.5 galaxies with a stellar mass ($M_{\star}$) $\geq$$10^{10}\,\msun$ are already quenched in both galaxy clusters ($>$50 per cent) and the field ($>$20 per cent), with clusters having a higher quenched fraction at all stellar masses compared to the field. A puzzling issue is that these massive quenched galaxies have stellar populations of similar age in both clusters and the field. This suggests that, despite the higher quenched fraction in clusters, the dominant quenching mechanism for massive galaxies is similar in both environments. In this work, we use data from the cosmological hydrodynamic simulations Hydrangea and EAGLE to test whether the excess quenched fraction of massive galaxies in $z$=1.5 clusters results from fundamental differences in their halo properties compared to the field. We find that (i) at $10^{10}\leq$$M_{\star}/\msun$$\leq 10^{11}$, quenched fractions at 1.5<$z$<3.5 are consistently higher for galaxies with higher peak maximum circular velocity of the dark matter halo ($v_{\mathrm{max, peak}}$), and (ii) the distribution of $v_{\mathrm{max, peak}}$ is strongly biased towards higher values for cluster satellites compared to the field centrals. Due to this difference in the halo properties of cluster and field galaxies, secular processes alone may account for (most of) the environmental excess of massive quenched galaxies in high-redshift (proto-)clusters. Taken at face value, our results challenge a fundamental assumption of popular quenching models, that clusters are assembled from an unbiased subset of infalling field galaxies. If confirmed, this would imply that such models must necessarily fail at high redshift, as indicated by recent observations.
\end{abstract}


%%%%%%%%%%%%%%%%%%%%%%%%%%%%%%%%%%%%%%%%%%%%%%%%%%

%%%%%%%%%%%%%%%%% BODY OF PAPER %%%%%%%%%%%%%%%%%%

\section{Introduction}
\label{sec:intro}
Understanding the quenching of star formation in galaxies as a function of mass and environment is a key unsolved question of contemporary astrophysics. A widely used model in the literature is the quenching model introduced by \citet{Peng2010} (``Peng model'' hereafter). Its key feature, motivated directly by $z<1$ observations, is that mass and environment affect the quenched fraction of galaxies in separable ways. This suggests the existence of two distinct quenching channels that are commonly referred to as mass- or self-quenching on the one hand, and environmental quenching on the other \citep{Peng2010,Peng2012}. Self-quenching depends on the galaxy stellar mass, plausibly through its connection with feedback from active galactic nuclei (AGN), which is likely to drive the quenching \citep[e.g.][]{tremonti2007,silk2012,fabian2012,Bower2017}. Environmental quenching, on the other hand, arises only in denser environments such as massive clusters of galaxies, where the interaction with other galaxies, including the brightest cluster galaxy (BCG), and with the hot gas in the host halo stop the star-formation process by stripping the galaxy of cold gas \cite[e.g.][]{gunn1972,larson1980,moore1998,boselli2022}.

The Peng model has the strength of using real observables such as the star formation rate (SFR) and stellar mass function (SMF) at $z\approx1$ to model the shape of the SMF of star-forming (SF) and quenched galaxies as a function of the environment in the local Universe, matching the Sloan Digital Sky Survey (SDSS) data. It also correctly describes the quenched fraction as a function of mass and environment. However, some recent studies have challenged the separability of mass and environment on quenched galaxy fractions at both low and high redshifts \citep[$0.3<z<3$; e.g.][]{darvish2016,pintoscastro2019,taylor2023}. 

Also, several studies have shown that this simple model fails to explain observations at high-redshift ($z>1$) dense environments \citep[e.g.][]{vdb2013,vanderburg2020GOGREEN}. In particular, \citet{vanderburg2020GOGREEN} studied a set of 11 massive clusters at $1 < z < 1.5$ from the Gemini Observations of Galaxies in Rich Early Environments (GOGREEN) project \citep{balogh2017,balogh2021} and found that the SMF of quenched galaxies has almost the same shape as in the (non-cluster) UltraVISTA field \citep{muzzin2013} at the same redshift. The \citet{Peng2010} model would predict them to look different, because the SMF of quenched galaxies in clusters is a product of both self-quenched and environmentally quenched galaxies, whereas field galaxies are only self-quenched. As long as there is any excess environmental quenching---which is clearly the case for the GOGREEN clusters---the shapes of quenched SMFs of cluster and field galaxies should therefore at no redshift look the same.

Moreover, \citet{webb2020} showed that the mean stellar ages of these quenched massive galaxies from the GOGREEN clusters are slightly older ($0.31^{+0.51}_{-0.33}$ Gyr at stellar mass range $10^{10}$ -- $10^{11.8} \msun$) than field galaxies (from UltraVISTA) at the same redshift, with an inferred quenching epoch at $z \gg 2$. This is firmly during the proto-cluster era, when the intracluster medium (ICM) was still very diffuse and is not expected to give rise to efficient ram-pressure stripping (RPS). {\modtextv If environmental quenching were significant for these massive cluster galaxies, one would expect them to have been quenched later than the field galaxies. In this case, cluster galaxies can potentially have younger stellar populations than field galaxies, contrary to what is observed. }

The small stellar age difference between quenched cluster and field galaxies observed by \citet{webb2020}, and the similar SMF of SF and quenched galaxies in cluster and field observed by e.g.~\citet{vdb2013} and \citet{vanderburg2020GOGREEN} hint at the necessity of an updated quenching model at high redshift that can explain both of the findings. One possibility is that galaxy quenching at high redshift is not only connected to the stellar mass but also the halo mass, similar to what is observed at lower redshift ($z\approx0.4$) by \citet{Mandelbaum2016}. If, at the same stellar mass, galaxies surrounded by a more massive halo are more likely to quench\footnote{The galaxy halo, or the halo surrounding a galaxy, is commonly referred to as the galaxy `subhalo' in simulations, unlike the group/cluster `halo' where the galaxy/subhalo resides.} \emph{and} these galaxies are proportionally more common in clusters, then massive quenched galaxies in both clusters and the field could quench purely as a result of their massive haloes. This would be independent of environmental quenching physics and could therefore happen even before they became cluster members, consistent with the observed ages \citep{webb2020}. A crucial prerequisite for this scenario is that clusters and the field at high-$z$ have different shapes of their halo mass functions (HMF). A similar suggestion was also made by \citet{werner2022} based on a higher satellite number density around massive quenched galaxies in the cluster infall region (for GOGREEN clusters, defined as the area within $1<r/r_{200}<3$ radial distance from the cluster centre) compared to quenched galaxies of similar stellar mass in the field at $z\approx1$. Their findings suggest that the infall region has a higher density of high-mass halos than the field, and this excess of massive haloes in the infall region can possibly enhance the galaxy quenching rate. 

One caveat to our above hypothesis is that at low redshift the shapes of the (sub-)halo mass function (HMF) in clusters and the field are identical \citep[e.g.][]{gao2012,bahe2017hydrangea}, although clusters have an (expected) normalisation offset in the HMF compared to the field due to the higher galaxy density. Based on these similar HMFs, the Peng model assumes that high density regions such as galaxy groups and clusters are simply accumulations of field galaxies. If cluster and field HMFs remain the same at high redshifts ($z>1$), this would imply that even in their formation epoch, galaxy clusters were just accumulations of field galaxies and that the increased number of quenched galaxies in clusters must be due to environmental quenching. However, if the shapes of the HMFs differ at $z \gtrsim 1$, then the excess quenched fraction in clusters can possibly be explained without any environmental quenching. Therefore, the HMFs at higher redshifts and in different environments need to be checked along with the quenched fractions for the stellar mass range of our interest. 

Such an investigation using only observational data is not possible, because it is not the current halo mass that matters (it is strongly affected by tidal stripping; see e.g.~\citealt{bahe2017hydrangea}), but the unobservable pre-infall (peak) halo mass. Therefore, the best approach to test this hypothesis is to use data from cosmological hydrodynamic simulations. In recent years, state-of-the-art cosmological simulations have been able to successfully reproduce many fundamental observable properties of different galaxy populations along with their DM halo masses, stellar mass functions (SMF), and density profiles across a significant fraction of the age of the Universe \citep[e.g.][]{dubois-horizon,schaye2014eagle,pillepich2018illustris,Dave_et_al_2019}. 

We use the Hydrangea suite of 24 zoom-in simulations of massive clusters \citep{bahe2017hydrangea} and the corresponding EAGLE 50 cMpc$^3$ volume \citep{schaye2014eagle} to test our hypothesis in this work. Both simulations were run using the exact same physics models, eliminating systematic offsets between the field and cluster galaxies for our analysis. The Hydrangea simulations successfully reproduce the observed galaxy SMF in clusters out to redshift 1.5 \citep{ahad2021}, and the EAGLE simulations reproduce observed field SMFs out to even higher redshifts considerably well \citep{Furlong2015eaglesmf}. Although the simulations fail to reproduce the quenched fraction of cluster galaxies at $z\sim1.5$, especially over-quenching the satellites at the low-mass end \citep{kukstas2023}, the most relevant parts for our analysis are the halo and stellar masses, which they do reproduce well. Furthermore, considering the use of the same galaxy formation model and a large enough galaxy sample size from both environments, EAGLE and Hydrangea are the most suitable companion simulations to test our hypothesis. {\modtextv However, we consider the potential impact of this mismatch of quenched fractions in our conclusions where relevant.}

The organization of the paper is as follows. In Sec.~\ref{sec:simdata}, we briefly introduce the simulations used in this work. We describe our sample selection and analysis in Sec.~\ref{sec:sample}. In Sec.~\ref{sec:result}, we present our results and discuss our interpretations in Sec.~\ref{sec:discussion}. Finally, we summarize our findings in Sec.~\ref{sec:conclusions}.

\section{Data}

\subsection{Simulations}
\label{sec:simdata}

We used data from the Hydrangea simulation suite \citep{bahe2017hydrangea, barnes2017cluster} for the clusters and data from the 50 Mpc$^3$ volume box of the EAGLE simulations \citep[see also \citealt{crain2015eagle}]{schaye2014eagle} for the field environment. The 50 Mpc$^3$ EAGLE simulations box used here was run with the `S15-AGNdT9' model \citep{schaye2014eagle}, which is the largest volume EAGLE box run with the exact same model as Hydrangea, ensuring a consistent comparison between clusters and field in our analysis. 

The Hydrangea simulations \citep{bahe2017hydrangea, barnes2017cluster}, part of the Cluster-EAGLE or `C-EAGLE' project, consist of high-resolution cosmological hydrodynamic zoom-in simulations of $24$ massive galaxy clusters. Each simulation region is centred on a massive cluster with $M_{200c}$ in the range $10^{14.0}$--$10^{15.4}\,\msun$ at $z = 0$\footnote{$M_{200c}$ refers to the mass within a sphere centred at the potential minimum of the cluster, and radius $r_{200c}$, within which the average density of matter is equal to 200 times the critical density.}. The high-resolution simulation boxes encompass $\geq 10$ virial radii ($r_{200c}$) of the cluster surroundings, making them also suitable to study the large-scale environmental influence on galaxies within and around clusters.

The resolution in both Hydrangea and the `S15-AGNdT9' EAGLE box are the same, with particle mass $m_\textrm{baryon} =1.81 \times 10^6\,\msun$ for baryons and $m_\textrm{DM} = 9.7 \times 10^6\,\msun$ for dark matter, respectively. The gravitational softening length is $\epsilon=0.7$ physical kpc (pkpc) at $z<2.8$. In both simulations, structures (galaxies and clusters) were identified in post-processing using the \subfind{} code \citep[see also \citealt{springel2001}]{dolag2009}, using a friends-of-friends (FoF) algorithm and subsequent identification of bound substructures.

A flat $\Lambda$CDM cosmology is assumed in both Hydrangea and EAGLE, with parameters taken from the \textit{Planck} 2013 results, combined with baryonic acoustic oscillations, polarization data from WMAP, and high multipole moment experiments \citep{planck2013}: Hubble parameter $H_0$ = 67.77 km s$^{-1}$ Mpc$^{-1}$, dark energy density parameter $\Omega_{\Lambda} = 0.693$, matter density parameter $\Omega_\textrm{M} = 0.307$, and baryon density parameter $\Omega_\textrm{b} = 0.04825$. 

\subsection{Sample selection}
\label{sec:sample}

{\modtextv We chose our dataset from all the available redshift snapshots from the simulations between redshifts 1.5 and 3.5, which were $\approx$1.5, 1.7, 2.0, 2.5, 3.0, and 3.5.} At all of these redshifts, our galaxy sample was selected from two distinct environments: massive clusters and the field. The cluster galaxies were chosen from the central clusters of each of the 24 Hydrangea zoom-in regions. All the subhaloes (or member galaxies) of the cluster from the FoF halo finder with a stellar mass of at least $10^{9}\,\msun$ and within the virial radius ($r_{200}$) of the corresponding cluster were included. To isolate the effects of environmental influence versus self-quenching, we also selected a subsample of the cluster galaxies by excluding the massive central galaxies and only keeping the satellites, which ensures that any galaxy in this subsample can be subjected to environmental quenching. This subsample is referred to as `cluster satellites' throughout this paper.

For our field galaxy sample, we selected all galaxies with $M_* \geq 10^{9}\msun$ from the EAGLE 50 Mpc$^3$ S15-AGNdT9 simulation. {\modtextv To separate the impact of environmental effects from self-quenching, we also selected the subset of central subhaloes from this field galaxy sample. This selection ensures that no galaxy in this subsample is subjected to environmental quenching, and therefore can only quench through internal processes. This sub-sample is referred to as `field centrals' throughout this paper.} 

\subsection{Galaxy properties}
\label{sec:gal_properties}
For the selected galaxies within each environment, we defined different galaxy properties based on the data from the simulation outputs. We selected the integrated mass of star particles within 30 physical kpc (pkpc) radius from the galaxy centre of potential as the galaxy stellar mass. In \citet{ahad2021}, we tested and verified that this definition of stellar mass is comparable to those obtained from running \textsc{SExtractor} \citep{bertin_arnouts1996} on the 2-dimensional projected stellar mass maps of the cluster galaxies. 

The galaxy halo mass was calculated from summing up the total mass of stars, dark matter, gas, and black hole particles that are connected to each subhalo. As the maximum circular velocity of the stars or gas is commonly used as an observational proxy to estimate the galaxy halo mass \citep[e.g.][]{nagai2005}, we also calculated the maximum circular velocity $v_{\mathrm{max}}$ for each galaxy, which we define as the maximum of $v = \sqrt{GM(<r)/r}$ where $r$ is the radial distance from the centre of potential of each galaxy and $M(<r)$ the total mass enclosed within a sphere of radius $r$. {\modtextv At each considered redshift from the simulations (as mentioned in Sec.~\ref{sec:sample})}, we also identified $v_{\mathrm{max, peak}}$, the maximum value of $v_{\mathrm{max}}$ for an individual galaxy across all previous snapshots including the current one. This is motivated by the findings of \citet{Reddick2013}, who show that $v_{\mathrm{max, peak}}$ of a dark matter halo correlates tightly with the properties of its galaxy. At each redshift, along with the $v_{\mathrm{max, peak}}$ for each galaxy, we selected the value of their stellar mass when $v_{\mathrm{max, peak}}$ occurred ($M_{\star}, v_{\mathrm{max, peak}}$ or $M_{\star,\,\mathrm{peak}}$), which can help with identifying the stellar mass growth at any redshift after $\vmaxpeak$ occurred through comparison with the stellar mass at that epoch. As an independent component from the $\vmaxpeak$, we also measured the peak halo mass, which is the maximum halo mass of an individual galaxy halo across all the redshifts from when the galaxy emerged until the redshift of interest.

We also use the spatially integrated star formation rate (SFR) for each galaxy in every snapshot, which we divide by $M_\star$ at the same redshift to obtain the specific SFR (sSFR). We consider galaxies with sSFR $\geq 10^{-10}$ yr$^{-1}$ as star-forming and those with lower sSFR as quenched. However, at $z\geq1$, the sSFR threshold separating star-forming galaxies from quenched ones is expected to evolve due to the evolution of their star-forming activity and stellar mass over cosmic time \citep[e.g.][]{Furlong2015eaglesmf}. We followed a similar principle as \citet{Furlong2015eaglesmf} and \citet{furlong2017} and applied an sSFR cut approximately one order of magnitude below the observed main sequence of star formation at each redshift of our concern. Our results in the following section are shown based on the fixed sSFR cut throughout this work because this threshold corresponds approximately to the separation between quenched and star-forming galaxies in \citet{kukstas2023} for $z = 1$.

\section{Results}
\label{sec:result}

\subsection{Halo mass function in cluster and field}
In this work, we test whether the quenching of massive galaxies in both cluster and field environments can be attributed to the distribution of their halo masses, rather than their stellar masses or environments. In the local Universe, the shape of the halo mass function of cluster galaxies is comparable to that in the field environment \citep[e.g.][]{bahe2017hydrangea}. At higher redshifts however, when the galaxy clusters are still forming, the halo mass distribution may well  be different between (proto-)clusters and the field. We therefore start by looking at the distribution of halo mass, or the halo mass function (HMF) in cluster and field at several redshifts between 1.5 and 3.5. We used the peak maximum circular velocity ($v_{\mathrm{max, peak}}$, see Sec.~\ref{sec:gal_properties} for details) of the galaxies to construct the HMF. 

%\begin{centering}
	\begin{figure*}
	\includegraphics[width=  1.9\columnwidth]{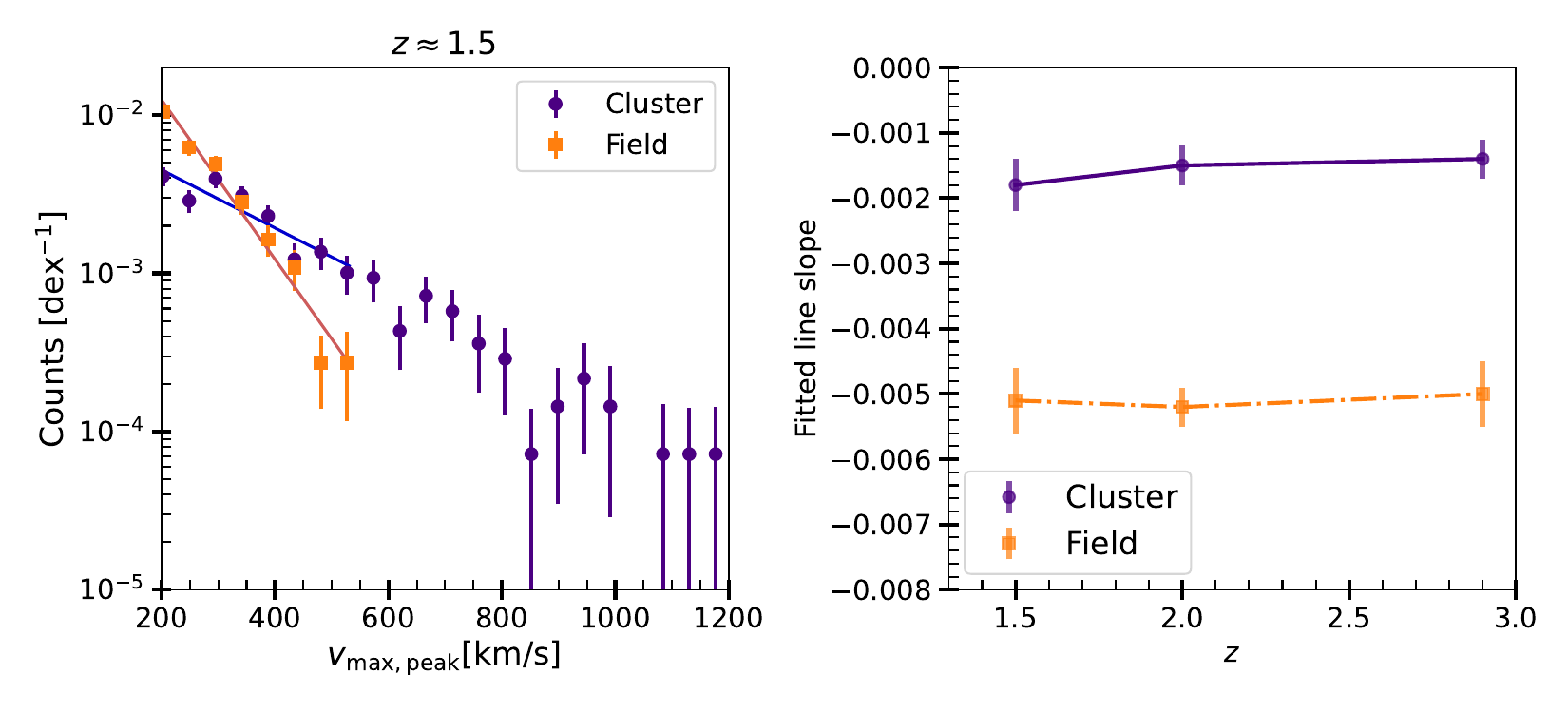}
	\caption{$v_{\mathrm{max,peak}}$ distribution in clusters (purple), and the field (orange) at $z=1.5$, normalized by the total number of galaxies in each of the environments. At the lower $v_{\mathrm{max,peak}}$ end, they are similar but they evolve differently from low to high $v_{\mathrm{max,peak}}$ in field and clusters, with clusters having a larger number of galaxies with high $v_{\mathrm{max,peak}}$ values. To demonstrate their different slopes, a log-linear function is fitted to the data points (shown by straight lines in left panel) and the slopes of the fitted lines are plotted in the right panel at different redshifts.}
	\label{fig:hmf_all}
	\end{figure*}
%\end{centering}

The HMF\footnote{We acknowledge that the distribution of $v_{\mathrm{max, peak}}$ is not exactly the distribution of the halo mass, i.e., the HMF, as we mention here. However, for their comparability and for simplifying the terms to compare with similar works, we use `HMF'.} at $z\approx1.5$ is shown in the left panel of Fig.~\ref{fig:hmf_all}. The HMFs here are normalised by the integrated values of the corresponding distributions and, consequently, represent the halo mass distribution if the field and cluster environments had the same number of galaxies. As the figure shows, compared to the field galaxies (orange points), galaxies in clusters (purple points) reach a much higher $v_{\mathrm{max, peak}}$. The highest values of $v_{\mathrm{max, peak}}$ in clusters are expected because clusters host, by definition, the most massive haloes. {\modtextv However, even at $v_{\mathrm{max, peak}}$ values where massive galaxies from both cluster and field exist ($v_{\mathrm{max, peak}} \geq 350 \mathrm{km/s}$), clusters have more galaxies than the field does.} This is more clearly demonstrated by the fitted lines in the distributions, as shown in Fig.~\ref{fig:hmf_all} with the same colour as the data points. The slope of the field HMF is steeper than the slope of the cluster HMF. The same behaviour is visible at all the redshifts of our consideration, shown in the right panel of Fig.~\ref{fig:hmf_all}. This panel shows that the slope of the field HMF is always steeper than the cluster HMF across our considered redshifts. Therefore, the HMFs in field and cluster environments are clearly different at all the redshifts we considered.

\subsection{Stellar-to-halo-mass relation and galaxy quenching}

To explore whether or how the different HMFs in different environments affect galaxy quenching, we study the stellar-to-halo-mass-relation (SHMR)\footnote{Similar to our use of the term `HMF' instead of `distribution of $v_{\mathrm{max, peak}}$', we use `SHMR' instead of `distribution of $v_{\mathrm{max, peak}}$ vs $M_*$' for their comparability and for simplifying the terms to compare with similar works.} of galaxies in our samples of clusters and field galaxies. For this part onward, we only select the `cluster satellite' and `field central' subsamples as described in Sec.~\ref{sec:sample}. This selection ensures that the field galaxies in consideration are not subject to any environmental quenching process and the cluster galaxies in our sample are exposed to the environmental quenching processes that are relevant at the corresponding redshifts. Similar to the HMF, we used the maximum circular velocity ($v_{\mathrm{max, peak}}$) of each galaxy halo in our sample at each environment to construct the SHMR. 

\begin{figure*}%[ht]
    \centering 
	\includegraphics[width=  1.9\columnwidth]{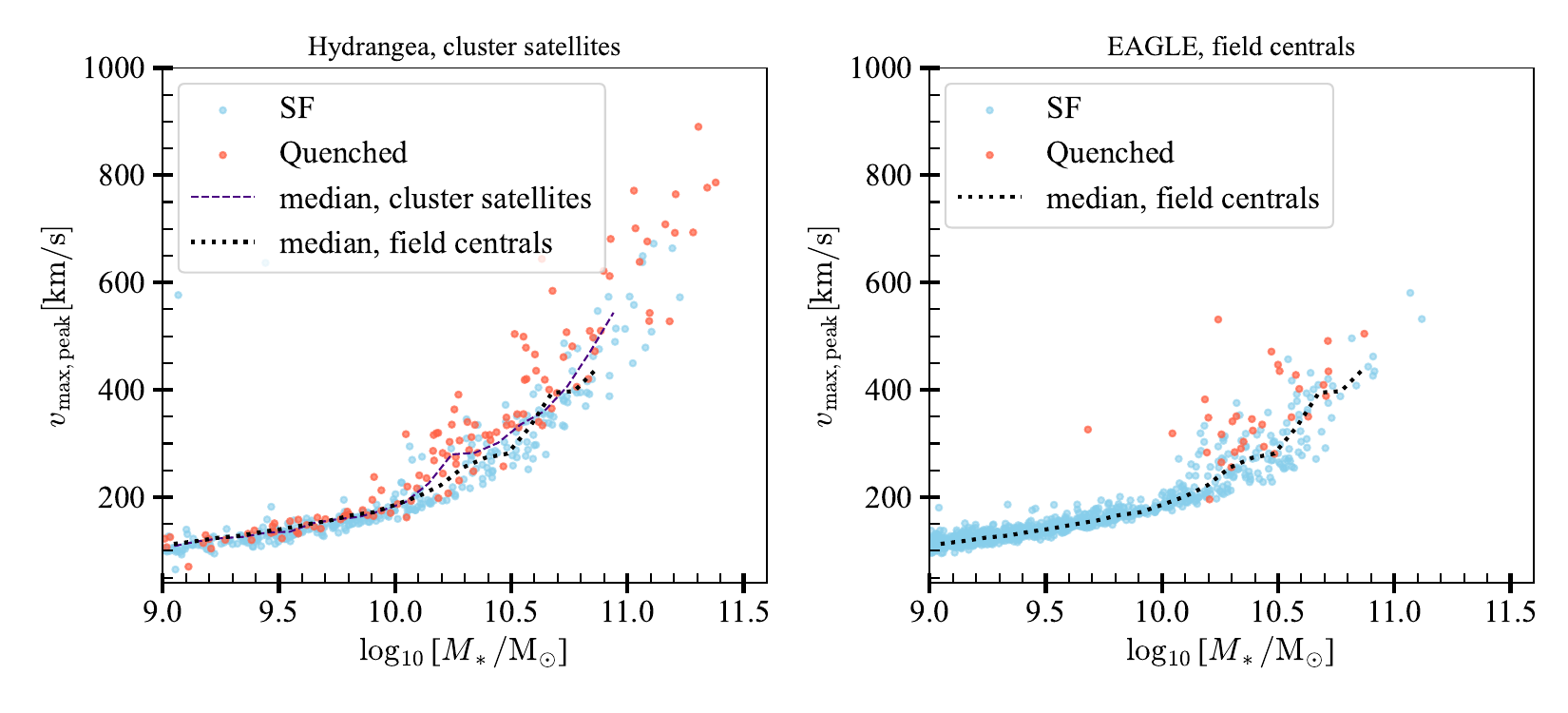}
	\caption{$v_{\mathrm{max, peak}}$ vs.~stellar mass (analogous to stellar-to-halo-mass) relations in clusters (left) and the field (right) at $z=1.5$. In both panels, red points represent galaxies that are quenched (sSFR$\leq$-10) while star-forming galaxies are shown in blue. The dashed (dotted) black lines indicate the running median distribution of the $v_{\mathrm{max, peak}}$ in cluster satellites (field centrals).}
	\label{fig:shmr_all}
\end{figure*}

Figure~\ref{fig:shmr_all} shows the SHMR in clusters (left) and the field (right) at $z=1.5$, respectively. In both panels, red points represent quenched (sSFR $\leq 10^{-10}$ yr$^{-1}$) galaxies and blue points represent star-forming ones (sSFR $> 10^{-10}$ yr$^{-1}$). The black lines indicate the running medians of the cluster (dashed) and field (dotted) samples. The dotted black line is also plotted in the left panel to allow a comparison of the running median values of both samples: at $M_*<10^{10}\,\msun$, the median $v_{\mathrm{max, peak}}$ values are similar in clusters and the field whereas above $10^{10}\,\msun$, the median $v_{\mathrm{max, peak}}$ values are higher in clusters. The same figure was constructed at the other redshifts considered (2.0 and 3.5); they show a similar characteristic as at $z=1.5$, albeit with a smaller fraction of quenched galaxies overall because they had less time to go through the quenching process. 

The smaller fraction of quenched galaxies at higher redshifts could also be partially due to our choice of a fixed sSFR to separate the quenched and star-forming galaxies at our redshifts of concern. Therefore, we reproduced the same figures with an evolving sSFR cut with redshift (as explained in Sec.~\ref{sec:gal_properties}). The exact number of quenched galaxies changed slightly with the evolving sSFR cut. However, our primary conclusions from this test with a fixed sSFR cut remained unchanged.

Three main features are visible in the panels of Fig.~\ref{fig:shmr_all}. First, there are more galaxies in clusters with higher stellar masses and $v_{\mathrm{max, peak}}$ than in the field, even when we only consider cluster satellites and field centrals. Second, for both environments, above $10^{10}\,\msun$ stellar mass, quenched galaxies (red) tend to have a higher $v_{\mathrm{max, peak}}$ (or halo mass) than star-forming galaxies (blue) at the the same stellar mass. Third, there are quite a few quenched galaxies below $10^{10}\,\msun$ stellar mass in clusters, whereas there are almost none in the field. All of these are interesting features to understand galaxy assembly and quenching, and require further investigation. However, before doing so, it is crucial to ensure that the second feature in Fig.~\ref{fig:shmr_all} is real and that the increased $v_{\mathrm{max, peak}}$ of the quenched galaxies does not occur after the quenching, rather than driving it.

%In both panels of Fig.~\ref{fig:shmr_all}, the quenched galaxies tend to have a higher $v_{\mathrm{max, peak}}$ value compared to the star-forming galaxies at the same stellar mass. 
There are two possible explanations for this feature: (i) There could be an upward shift in $v_{\mathrm{max, peak}}$ between the quenched and star-forming galaxies at fixed stellar mass, which would imply that the quenched galaxies indeed have higher $v_{\mathrm{max, peak}}$ (or higher halo-mass). Alternatively (ii), there could be a shift to the left in stellar mass (i.e.~to lower masses) of quenched galaxies at a fixed $v_{\mathrm{max, peak}}$, which implies that the quenched galaxies with the same $v_{\mathrm{max, peak}}$ did not grow as much in stellar mass compared to the star-forming ones. The upward-shift scenario occurs when $v_{\mathrm{max, peak}}$ rises either before or after the galaxy quenches. The left-shift scenario, on the other hand, can occur either due to significant stellar mass stripping of quenched galaxies, or because stellar mass growth stops for quenched galaxies while star-forming galaxies continue to grow and hence move to the right in Fig.~\ref{fig:shmr_all}).

%Discuss how the top-left shift of quenched galaxies can be for either shifting of the points towards the top (increase in vmax), left-shifting of the galaxies (significant stripping of stars), or lack of right shifting of quenched galaxies (SF galaxies will keep forming stars and hence move towards right). 

\begin{center}
	\begin{figure*}
		\includegraphics[width=  2\columnwidth]{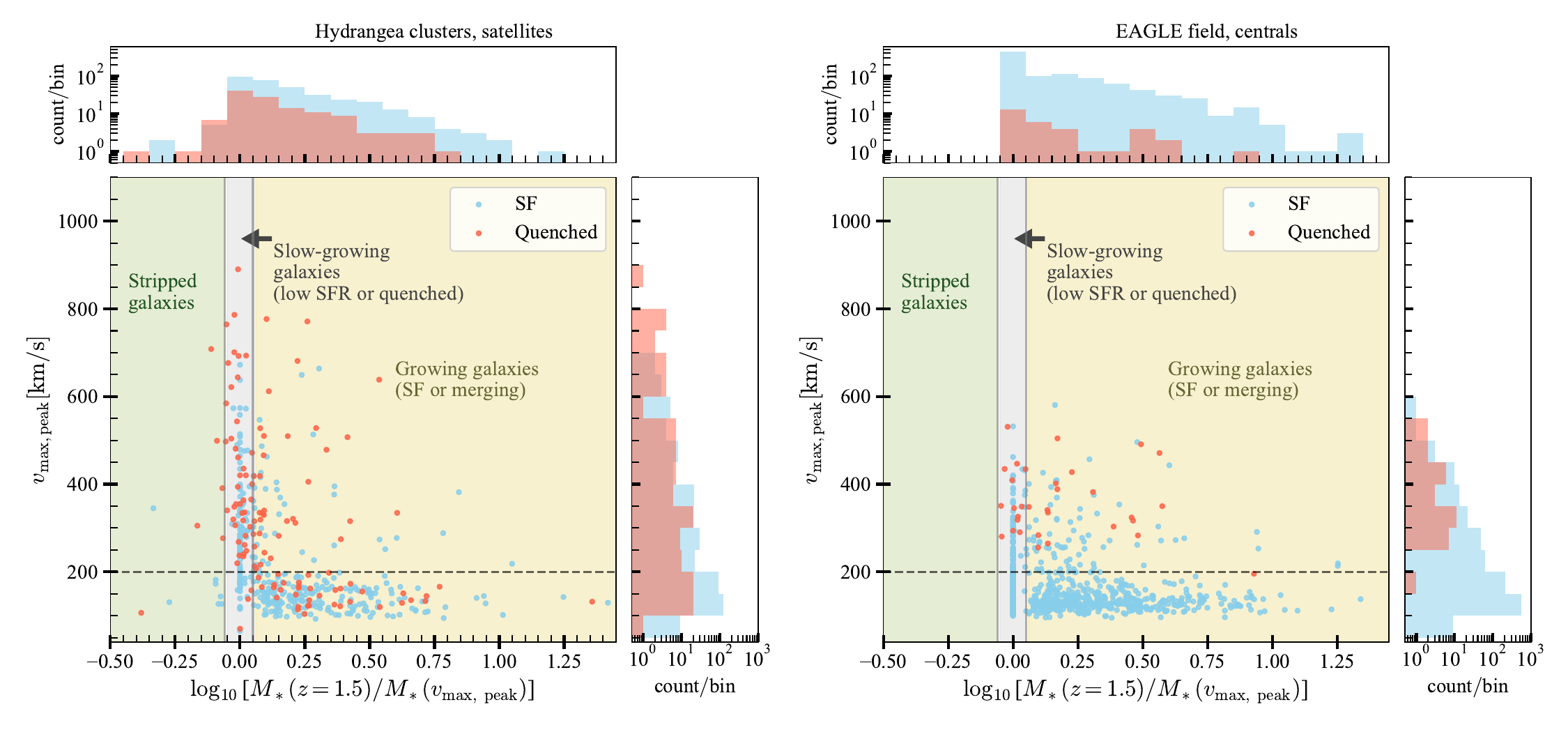}
		\caption{$v_{\mathrm{max, peak}}$ vs stellar mass ratio in clusters (left) and the field (right). The stellar mass ratio is obtained from the ratio of the galaxy stellar mass at $z=1.5$ and the galaxy stellar mass at the epoch when $v_{\mathrm{max, peak}}$ occurs. In both panels, the colours of the data points indicate whether they are quenched (sSFR$\leq$-10, red) or star-forming (blue). {\modtextv Also, in both panels, the histograms on top and right show the distribution of galaxies along the projected axis. The horizontal dashed line at 200~km/s separates the low-mass galaxies (below) from the high-mass galaxies (above)}. The ratio of stellar mass at $z=1.5$ and stellar mass where $v_{\mathrm{max, peak}}$ occurs indicates what happened to the galaxy since $v_{\mathrm{max, peak}}$ occurred. If the galaxy went through stripping of stars since $v_{\mathrm{max, peak}}$ occurred, it will have a negative log mass ratio, as indicated by the light green area in both panels. If galaxies kept on growing since $v_{\mathrm{max, peak}}$ occurred (primarily by star formation or merging), they would have a positive log mass ratio, indicated by the yellow area. If a galaxy stopped forming stars and did not grow in stellar mass since $v_{\mathrm{max, peak}}$, it will be at or near the 0 log mass ratio location, shown by the grey area. This last case can also occur if $v_{\mathrm{max, peak}}$ happened at or close to $z=1.5$.}
		\label{fig:ms_mspeak_ratio}
	\end{figure*}
\end{center}

To identify which of these two scenarios is the dominant reason for the separation of the star-forming and quenched galaxies in Fig.~\ref{fig:shmr_all}, we took the stellar masses of each galaxy at the redshift when its $v_{\mathrm{max, peak}}$ occurs ($\mstarpeak$), and plotted the $v_{\mathrm{max, peak}}$ vs $\mstar (z=1.5) / \mstarpeak$ relation in Fig.~\ref{fig:ms_mspeak_ratio}. This ratio of stellar mass at $z=1.5$ to the stellar mass at $\vmaxpeak$ is an indication of how the stellar mass of the galaxy has changed since the epoch of $\vmaxpeak$. There are three possible areas where galaxies can be in this plot (shown by shaded regions in Fig.~\ref{fig:ms_mspeak_ratio}), and each location implies a possible scenario about why they are there. (i) If significant stellar stripping were responsible for the left-shift of quenched galaxies in Fig.~\ref{fig:shmr_all}, then $\mstar (z=1.5) < \mstarpeak$ and hence $\log(\mstar (z=1.5) / \mstarpeak) < 0$. (ii) If quenched galaxies have $\mstar (z=1.5) \approx \mstarpeak$ whereas star-forming galaxies have $\mstar (z=1.5) \gg \mstarpeak$, it would imply that a strong right-shift of star-forming galaxies (and lack thereof for quenched galaxies) created the trend visible in Fig.~\ref{fig:shmr_all}. If $\mstar (z=1.5) \approx \mstarpeak$ for both star-forming and quenched galaxies, it could also imply that $v_{\mathrm{max, peak}}$ occurred recently in both environments. And finally, (iii) if there is no substantial difference in the distribution of $\log\,(\mstar (z=1.5) / \mstarpeak)$ values between star-forming and quenched galaxies, and both span a broad range in the positive $x-$axis, it would imply that the primary cause of the separation of the star-forming and quenched galaxies in Fig.~\ref{fig:shmr_all} is not a horizontal shift of the galaxies along the stellar mass axis. 

Figure~\ref{fig:ms_mspeak_ratio} shows that at redshift 1.5, only a handful of cluster satellites are in the area that shows stripped galaxies (light green shaded region), i.e., only a few show any sign of strong stellar stripping. In addition, not all of these galaxies with $\mstar (z=1.5) < \mstarpeak$ are quenched. Therefore, this scenario can be ruled out as the primary reason for the star-forming and quenched galaxy separation in Fig.~\ref{fig:shmr_all}. Figure~\ref{fig:ms_mspeak_ratio} also shows that in both environments, {\modtextv a good fraction of both star-forming and quenched galaxies\footnote{\modtextv In the narrow grey shaded region in Fig.~\ref{fig:ms_mspeak_ratio}, where $\mstar (z=1.5)$ is close to the value of $\mstarpeak$, there are about 28 per cent and 46 per cent of the star-forming galaxies from clusters and the field, respectively. As for the quenched galaxies in the same area, the amount is about 36 per cent in clusters and 41 per cent in the field. However, if we only consider galaxies that acquire their $\vmaxpeak$ exactly at $z=1.5$, the quantities go sharply down to about 2 per cent in clusters and 3 per cent in the field for quenched galaxies, and 18 per cent in clusters and 45 per cent in the field for the star-forming galaxies.} have $\mstar (z=1.5) \approx \mstarpeak$ (i.e.~close to 0 along the $x$-axis)}, whereas the rest of them scatter up to $\log\,(\mstar (z=1.5) / \mstarpeak) \approx 1$, with no clear separation between star-forming and quenched galaxies. This implies that a right-shift of star-forming galaxies alone cannot explain the separation of star-forming and quenched galaxies in Fig.~\ref{fig:shmr_all}. Although we only show $z = 1.5$ in Fig.~\ref{fig:ms_mspeak_ratio}, we have verified that the same conclusion holds at the other considered redshifts between 1.5 and 3.5.
%Describe how this plot cancels the left-shifting of the galaxies (significant stripping of stars), and lack of right shifting of quenched galaxies (SF galaxies will keep forming stars and hence move towards right) scenarios. The primary reason why the quenched galaxies are at the top-left of the SF galaxies is because of the upshift of Vmax-peak.
%\section{Discussion}

{\modtextv In short, the best explanation for the offset of quenched galaxies to higher $\vmaxpeak$ values at fixed $\mstar > 10^{10}\,\msun$ as seen in Fig.~\ref{fig:shmr_all} is either the stellar-mass growth of star-forming galaxies, or an intrinsically deeper potential well, or a combination of these two. The more interesting one for this work is the case where a deeper potential well is responsible for this separation, which is also likely to be connected to their quenching mechanism, e.g. AGN feedback.}

\subsection{Quenched fraction and halo mass}

\begin{center}
	\begin{figure*}
		\includegraphics[width=  1.9\columnwidth]{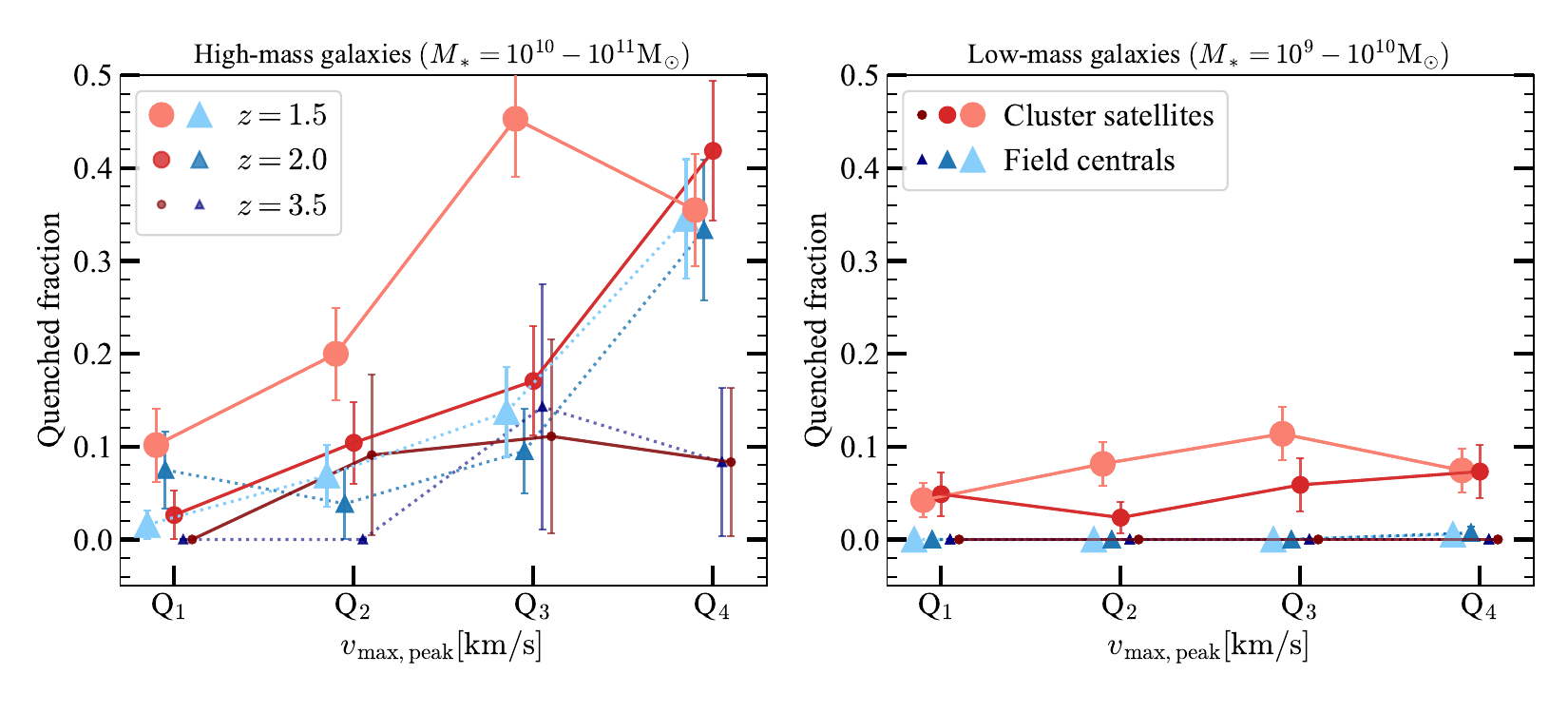}
		\caption{Fraction of quenched galaxies vs.~the $v_{\mathrm{max, peak}}$ quartile in clusters (red shades) and field (blue shades) at different redshifts. The evolution is shown for two different galaxy mass ranges, $10^{10} - 10^{11} \msun$ (left-hand panel) and $10^9 - 10^{10}\msun$ (right-hand panel). Higher redshift data points are respectively smaller and in darker tones or the same colours. There is no variation of quenched fractions with $v_{\mathrm{max, peak}}$ quartile for low-mass galaxies. However, for the high-mass galaxies, a higher $v_{\mathrm{max, peak}}$ corresponds to a higher quenched fraction at $z\leq2$. This trend is visible both in cluster satellites and field centrals.}
		\label{fig:qf_vs_vmax_range}
	\end{figure*}
\end{center}

\noindent %The remaining scenario to test is the relation between galaxy quenching and a higher $v_{\mathrm{max, peak}}$. 
To demonstrate the correlation between quenching and $\vmaxpeak$ more clearly, we show in Fig.~\ref{fig:qf_vs_vmax_range} the fraction of quenched cluster and field galaxies in running quartiles of $v_{\mathrm{max, peak}}$. The quenched fraction here is measured in running $v_{\mathrm{max, peak}}$ quartiles\footnote{One of the $v_{\mathrm{max, peak}}$ bin separators, the running median, is shown by the black dashed (clusters) and dotted (field) lines in Fig.~\ref{fig:shmr_all}. The other two are the running 25th and 75th percentiles.}. % by dividing the total number of quenched galaxies (red points in Fig.~\ref{fig:shmr_all}) by the total number of galaxies . 
The relation for clusters is shown by circles in red shades and for the field by blue shaded upward triangles. In each case, we show three different redshifts, with higher redshifts represented by smaller markers and darker shades. For clarity, the trends are shown separately for two broad stellar mass bins, low-mass galaxies with $\mstar$ between $10^9$ and $10^{10}\,\msun$ in the right-hand panel, and high-mass galaxies ($\mstar$ between $10^{10}$ and $10^{11}\,\msun$) on the left.

A number of key features are worth pointing out in Fig.~\ref{fig:qf_vs_vmax_range}. Starting with the low-mass field galaxies (blue triangles in the right-hand panel), we see that there are almost no quenched galaxies at any redshift or $v_{\mathrm{max, peak}}$. For low-mass galaxies in clusters (red circles in the right-hand panel), the same is true at the highest redshift in this figure, $z = 3.5$. At $z\approx2.0$, however, we see that about 5 per cent of low-mass galaxies are already quenched in all four $v_{\mathrm{max, peak}}$ quartiles. At $z = 1.5$, the fraction of quenched low-mass galaxies increases to approximately 9 per cent in the $2^\mathrm{nd}$ and $3^\mathrm{rd}$ $v_{\mathrm{max, peak}}$ quartiles, while remaining unchanged in the $1^\mathrm{st}$ and $4^\mathrm{th}$ quartiles. In other words, there is no dependency of quenched fraction on $v_{\mathrm{max, peak}}$ for low-mass galaxies in clusters or the field, but low-mass cluster galaxies are beginning to quench due to environmental mechanisms\footnote{Instead of a constant sSFR threshold across redshifts, we repeated this analysis using an evolving sSFR threshold with redshift, e.g.~a fixed offset from the star-forming main sequence to define quenched galaxies. There is a slight hint that the first signs of environmental quenching appear even earlier than shown in Fig.~\ref{fig:qf_vs_vmax_range}, but this change of the sSFR threshold did not affect our primary conclusions.} unrelated to $\vmaxpeak$ by $z = 2$.

%Also, at $z\approx3.5$, low-mass galaxies in the field and clusters have the same quenched fractions, which increase in clusters uniformly across all $v_{\mathrm{max, peak}}$ bins at lower redshifts. As the $v_{\mathrm{max, peak}}$ distribution for low mass galaxies is similar in clusters and the field (Fig.~\ref{fig:shmr_all}), the increased quenched fraction must result from environmental quenching in clusters which became important at this stellar mass somewhere between redshifts 3.5 and 2.0. 

Considering instead the high-mass galaxies (left-hand panel in Fig.~\ref{fig:qf_vs_vmax_range}), we see that, in both clusters and the field, the quenched galaxy fraction correlates almost always with $v_{\mathrm{max, peak}}$. This behaviour is consistent with our hypothesis that, at fixed stellar mass, halo properties (such as a higher halo mass) can make galaxies more likely to be quenched irrespective of the environment they reside in. {\modtextv We also re-constructed Fig.~\ref{fig:qf_vs_vmax_range} by excluding the galaxies that acquire their $\vmaxpeak$ exactly at $z=1.5$ (not shown here). Given the small fraction of quenched galaxies that meet this condition (see Footnote~5), this selection only makes a small upward shift of the trend of quenched fractions versus the $\vmaxpeak$ quartiles, but does not affect the main conclusions from this work.}%However, unlike the low-mass galaxies, where the effect of environment is clearly separable from self-quenching (as there are no quenched galaxies among the field centrals at this mass range), for high-mass galaxies, the effect of environment and self quenching (correlation with higher halo mass) is entangled as both effects are at play in cluster satellites. (check the wording)

\section{Discussion}
\label{sec:discussion}

\subsection{Quenched fraction in clusters and the field}

To understand the effects of environmental and self-quenching in our high-mass galaxy samples from Fig.~\ref{fig:qf_vs_vmax_range}, we need to compare the quenched fractions in clusters (red shades) and the field (blue shades) at the same redshifts (similar marker size as indicated in the labels) in the left panel of Fig.~\ref{fig:qf_vs_vmax_range}. At $z = 3.5$, the quenched fractions of cluster and field galaxies are similar for both the high- and low-mass galaxy samples, implying that environmental quenching was negligible at very high redshift, regardless of galaxy mass. At $z = 2$, the quenched fraction of high-mass cluster satellites is still relatively comparable (with a slight increase) to the high-mass field centrals (medium dark red and blue symbols in Fig.~\ref{fig:qf_vs_vmax_range}, left-hand panel). The environmental effect is similarly low for the low-mass galaxies, hinting that at $z = 2$, the environmental effect in clusters is already present, albeit weakly. At $z=1.5$, the difference in the quenched fraction is prominent for high-mass galaxies. Especially in the lowest three $v_{\mathrm{max, peak}}$ quartiles, the quenched fractions in clusters are almost a factor of two higher at the same $v_{\mathrm{max, peak}}$ compared to the field, which is likely due to environmental effects. Interestingly, at the highest $v_{\mathrm{max, peak}}$ bin, the quenched fraction is almost the same in clusters and the field. The high-mass quenched fractions in the field increase only marginally between $z = 2$ and 1.5, by $< 0.05$. %Comparing the high-mass quenched fractions in the field between $z = 2$ and 1.5, there is almost no evolution there as well. 

The most plausible explanation is that, for massive galaxies, self-quenching is directly correlated to $v_{\mathrm{max, peak}}$ while environmental quenching is either inversely correlated or uncorrelated with $v_{\mathrm{max, peak}}$. At the same high stellar mass range, if a galaxy has a higher $v_{\mathrm{max, peak}}$, it is similarly likely to be quenched in both clusters and the field, but for lower $v_{\mathrm{max, peak}}$ ranges, it is more likely to be quenched as a satellite in a massive cluster compared to being an isolated or central galaxy in the field. This is likely because, if a galaxy in a dense environment has a higher halo mass, it will retain its cold gas more efficiently than a galaxy with the same stellar mass but a smaller halo mass, resulting in slower or less efficient quenching. Therefore, at this stellar mass range ($10^{10} - 10^{11} \msun$), in redshift 1.5 clusters, along with galaxies that quenched after becoming a satellite, there can also be satellites that were already quenched in the field before being accreted onto the cluster. In this scenario, the age of the stellar population in quenched cluster galaxies would be comparable to that of similarly massive quenched field galaxies, consistent with the results from the GOGREEN survey \citep{webb2020}.

{\modtextv One point of concern for our finding is that the EAGLE and Hydrangea simulations were found to have a mass-dependent mismatch to the observed quenched galaxy fractions at $z>1$ \citep{kukstas2023}. The field galaxies have a good match to the quenched fractions (QF) out to $10^{10.7} \msun$, but they are under-quenched above this stellar mass limit. In the Hydrangea clusters, the discrepancy is at both lower and higher mass ends, with over-quenching below $10^{9.8} \msun$ and slight under-quenching above $10^{10.8} \msun$. If the QF is significantly different between observations and our tested simulations at the concerned mass range, then the impact of such a mismatch on our conclusions needs to be adjusted accordingly. 

As for the over-quenching of low-mass satellites ($< 10^{9.8} \msun$) in \citet{kukstas2023}, the mass range matches our low-mass galaxy sample. We do not see any trend with respect to the halo properties for such galaxies. Also, in the field environment (EAGLE simulations), where the quenched fractions match the observations quite well in this mass range, we see that the trend of no dependence of quenched fractions with respect to halo properties holds true here as well. Therefore, the over-quenching of low-mass satellites is unlikely to affect our conclusions significantly.

For the high-mass galaxies ($> 10^{10} \msun$) in Hydrangea simulations, the right panel of fig.~6 from \citet{kukstas2023} shows that the observed QF overlaps within errorbars out to $10^{10.8} \msun$. Only about 11 per cent of our high-mass galaxy sample between $10^{10}$ and $10^{11} \msun$ occupy this slightly under-quenched region. Given that most of our high-mass galaxies show a reasonable match with the observed QF, we conclude that it is also unlikely that the mismatch of QF at the high-mass end significantly affects our conclusions from this work.}

\subsection{Do $v_{\mathrm{max, peak}}$ and halo mass have the same effect on quenching?}

In our analysis, we have chosen $v_{\mathrm{max,\, peak}}$ as proxy for the halo mass of a galaxy, motivated by its similarity to observational methods. However, the simulated halo catalogue also provides (galaxy) halo masses directly, as well as  their peak values across cosmic time. We used %two different definitions of the peak halo masses of our galaxy samples (
the peak halo mass of our galaxy sample across the considered redshifts, and %halo mass at the epoch when $v_{\mathrm{max,\, peak}}$ occurs, details in Sec.~\ref{sec:gal_properties}) and
repeated the same analysis with the HMF, SHMR, and quenched fraction of galaxies in halo mass quartiles. We found somewhat similar results for the halo mass (more details in Appendix~\ref{sec:app_qf_mh}). However, compared to the $v_{\mathrm{max, peak}}$, the findings in terms of the peak halo mass have more scatter. The most likely reason for this is that quenching is actually controlled by a third property that correlates with both halo mass and $v_{\mathrm{max,\,peak}}$, the latter correlation being stronger. 

One promising candidate that drives the quenching more directly is the mass of the central supermassive black hole (SMBH, $M_{\mathrm{SMBH}}$). A high $M_{\mathrm{SMBH}}$ indicates high past accretion and AGN feedback activity of the SMBH, which can drive the quenching, as shown by multiple simulation- and observation-based works \citep[see e.g.][]{Bower2017,piotrowska2022}. Besides, the size (half-mass radius, $r_{\mathrm{1/2, mass}}$) of the galaxies is a likely parameter that adds to the scatter between halo mass and $v_{\mathrm{max,\,peak}}$. Due to the definition of $v_{\mathrm{max, peak}}$ (see Sec. \ref{sec:gal_properties}), a low  $r_{\mathrm{1/2, mass}}$ can increase the value of $v_{\mathrm{max, peak}}$ since it indicates a higher concentration of mass near the galactic centre. Furthermore, if a galaxy came too close to the cluster centre not long before reaching its $v_{\mathrm{max, peak}}$, the quenching is more likely to be environmental, irrespective of the $v_{\mathrm{max, peak}}$ value. Therefore, along with $r_{\mathrm{1/2, mass}}$ and $M_{\mathrm{SMBH}}$, we also tested the correlation of the minimum distance from the cluster centre reached by each galaxy before they attained their $v_{\mathrm{max, peak}}$ to the quenched galaxies in our high-mass galaxy sample. 

Only 27 per cent of the quenched cluster satellites showed a close proximity (within 20 per cent of their instantaneous virial radius) to the cluster centre before reaching their $\vmaxpeak$, excluding the majority of these early-quenched galaxies (more details in Appendix~\ref{sec:app_mbh_r_half}). Therefore, most of the cluster satellites in our sample were not close enough to the cluster centre to experience environmental quenching via strong tidal or hydrodynamic forces before reaching their $\vmaxpeak$. 

{\modtextv In field centrals, where self-quenching must be the dominant mechanism, 85 per cent of quenched galaxies have central black hole masses above $10^{7.5}~\msun$\footnote{This mass limit was arbitrarily chosen to keep all the quenched field centrals with a massive black hole above the threshold. See top right panel of Fig.~\ref{fig:mbh_r_half_vmax}.}. Similarly, 41 per cent of quenched cluster satellites have a high black hole mass (above $10^{7.5}~\msun$), which indicates that at least for 41 per cent of the quenched cluster galaxies, AGN feedback can be the dominant quenching mechanism. In comparison, about 24 per cent of the star-forming field centrals and about 18 per cent of the star-forming cluster satellites have central black hole masses above $10^{7.5}~\msun$, which is less than half compared to the quenched galaxies in the same environments.} As far as their sizes are concerned, approximately 70 per cent of quenched galaxies have a small half-mass radius ($\leq 1$ kpc)  in both clusters and the field. More details on this test are provided in Appendix~\ref{sec:app_mbh_r_half}.

Although most of the quenched galaxies have either a massive central black hole or a small size (or both, for a few), none of these properties can completely explain the scatter between the effects of halo mass and $v_{\mathrm{max, peak}}$ on galaxy quenching. Previous studies have found that, instead of the galaxy mass, smaller galaxy size ($<1$ kpc) and higher central density can also be connected to inducing galaxy quenching, and these properties are connected to the central baryonic properties of the galaxy \citep[e.g.][]{yano2016,whitaker2017}. These findings hint that the central baryonic properties of a galaxy could be responsible for the scatter that we find between how halo mass or $\vmaxpeak$ affect quenching. Instead of halo mass, $v_{\mathrm{max, peak}}$ better captures this connection between the central baryonic properties of a galaxy and the quenching of its star formation. Models of galaxy quenching at high redshifts should therefore consider their halo mass and $v_{\mathrm{max, peak}}$, along with their stellar mass and environment.

\section{Summary and Conclusions}
\label{sec:conclusions}

In recent works based on the GOGREEN cluster survey, a similar shape of the stellar mass function of quenched galaxies in clusters and the field \citep{vanderburg2020GOGREEN}, and similar ages of massive quenched galaxies in both environments \citep{webb2020}, have been observed. These findings cannot be explained using the widely-accepted model for separable mass- and environmental quenching proposed by \citet{Peng2010}, and therefore necessitate a review of existing galaxy quenching models, especially at high redshifts. In this work, we tested whether a difference in the halo mass function between cluster and field environments above redshift 1.5, along with a halo-mass dependant quenching efficiency, can explain the observed discrepancies by using data from the cosmological hydrodynamic simulations Hydrangea and EAGLE. Our findings are as follows:

\begin{enumerate}
    \item The normalized distribution of $v_{\mathrm{max, peak}}$ has a different shape in cluster and field environments at $z \geq 1.5$. The shape of the distributions, quantified by the slope of a fitted line, was consistently different between clusters and the field out to redshift 3 (Fig.~\ref{fig:hmf_all}).

    \item The stellar mass to $v_{\mathrm{max, peak}}$ relations of cluster satellites and field centrals show that {\modtextv most quenched galaxies have a $v_{\mathrm{max, peak}}$ value higher than the median $v_{\mathrm{max, peak}}$ at any given stellar mass above $10^{10}\msun$ (Fig.~\ref{fig:shmr_all})}. This behaviour suggests that a higher $v_{\mathrm{max, peak}}$ (which is similar to observational proxies of halo mass) may be correlated with the quenching of high-mass galaxies above redshift 1.5.

    \item We see almost no quenched galaxies among $z= 3.5$ low mass galaxies {\modtextv ($9<\log(M_*/\msun)<10$)} in both field and clusters, which remains the same in low-mass field centrals even at $z= 1.5$. Low-mass cluster satellites however, show an increase in their quenched fraction over time, independent of  $v_{\mathrm{max, peak}}$, which grows to 10 per cent by redshift 1.5. This increase demonstrates that environmental quenching becomes significant between $2<z<3.5$ in these low-mass galaxies (Fig.~\ref{fig:qf_vs_vmax_range}). The absence of quenched galaxies in this stellar mass range for field centrals also indicates that environmental quenching is the primary mechanism for quenching these low-mass galaxies.

    \item High-mass galaxies {\modtextv ($10<\log(M_*/\msun)<11$)} in both clusters and the field have a clear correlation of their quenched fractions with $v_{\mathrm{max, peak}}$ at all redshifts considered in this work ($1.5 \leq z \leq 3.5$). This suggests that the same mechanism(s) quenched high-mass field centrals and cluster satellites, at least until redshift 2. 
    
    \item For high-mass galaxies, the enhancement of quenched fraction is higher in cluster satellites compared to the field centrals between $z=$ 2 and 1.5, indicating a contribution from environmental quenching in high-mass cluster satellites. However, for the highest $v_{\mathrm{max, peak}}$ quartile, the quenched galaxy fraction is comparable in field and clusters at all the considered redshifts, suggesting that at fixed stellar mass, galaxies with the highest halo mass are the least affected by the environmental quenching compared to galaxies with a lower halo mass. 

\end{enumerate}

Our finding is qualitatively consistent with the discussions of \citet{vanderburg2020GOGREEN}, that the stellar mass functions of quenched galaxies in clusters and field at $z= 1.5$ have a similar shape because their primary quenching mechanisms were similar. Summary point (v) is also qualitatively consistent with the findings of \citet{webb2020}, explaining that at $z=1.5$, high-mass cluster satellites can be both environmentally quenched or self-quenched before they enter into the clusters and, therefore, can have a comparable age of the stellar population to the quenched field galaxies.

Compared to the existing simple quenching models that separate mass and environmental quenching based on low-redshift observations, this work using cosmological simulations better reconciles the high quenched fraction in clusters with the lack of an age dependence of quiescent galaxies on the environment. If true, it also implies that the majority of high-mass galaxies in (proto-)clusters are quenched by secular processes, not by their environment. 

Finally, the differing halo mass functions imply that (proto-) clusters do not grow simply from the infall of field halos. Therefore, galaxy quenching models at high redshifts need careful revision, especially with improved observations and insights in the {\it JWST} era. Considering the effect of the underlying halo-mass distributions and central baryonic concentrations will be a valuable starting point in this direction.

\section*{Acknowledgements}

{\modtextv We thank the reviewer for valuable comments that helped to improve the presentation of our results.} The authors acknowledge support from the Netherlands Organization for Scientific Research (NWO) under Vici grant number 639.043.512 (SLA, HH), and Veni grant number 639.041.751 (YMB). YMB also gratefully acknowledges financial support from the Swiss National Science Foundation (SNSF) under project 200021\_213076. The Hydrangea simulations were in part performed on the German federal maximum performance computer ``HazelHen'' at the maximum performance computing centre Stuttgart (HLRS), under project GCS-HYDA / ID 44067 financed through the large-scale project ``Hydrangea'' of the Gauss Center for Supercomputing. Further simulations were performed at the Max Planck Computing and Data Facility in Garching, Germany. This work used the DiRAC@Durham facility managed by the Institute for Computational Cosmology on behalf of the STFC DiRAC HPC Facility (www.dirac.ac.uk). The equipment was funded by BEIS capital funding via STFC capital grants ST/K00042X/1, ST/P002293/1, ST/R002371/1 and ST/S002502/1, Durham University and STFC operations grant ST/R000832/1. DiRAC is part of the National e-Infrastructure.

The analysis of this work was done using Python (\href{http://www.python.org}{http://www.python.org}), including the packages \textsc{NumPy} \citep{Harris_et_al_2020}, \textsc{AstroPy} \citep{astropy2013}, and \textsc{SciPy}
\citep{jones2009}. Plots have been produced with \textsc{Matplotlib} \citep{hunter2007matplotlib}. 

%%%%%%%%%%%%%%%%%%%%%%%%%%%%%%%%%%%%%%%%%%%%%%%%%%
\section*{Data Availability}

The data presented in the figures are available upon request from the corresponding author. The Hydrangea data are available at \href{https://ftp.strw.leidenuniv.nl/bahe/Hydrangea/}{https://ftp.strw.leidenuniv.nl/bahe/Hydrangea/}. The Eagle simulations are publicly available; see \citet{mcalpine2016,eagle2017} for how to access EAGLE data.



\bibliographystyle{mnras}
\bibliography{high_z_GQ} % if your bibtex file is called example.bib

%\begin{thebibliography}{99}
%\end{thebibliography}

%%%%%%%%%%%%%%%%%%%%%%%%%%%%%%%%%%%%%%%%%%%%%%%%%%

%%%%%%%%%%%%%%%%% APPENDICES %%%%%%%%%%%%%%%%%%%%%

\appendix

\section{Quenched fraction vs. halo mass quartile}
\label{sec:app_qf_mh}

\begin{center}
    \begin{figure}
        \includegraphics[width= \columnwidth]{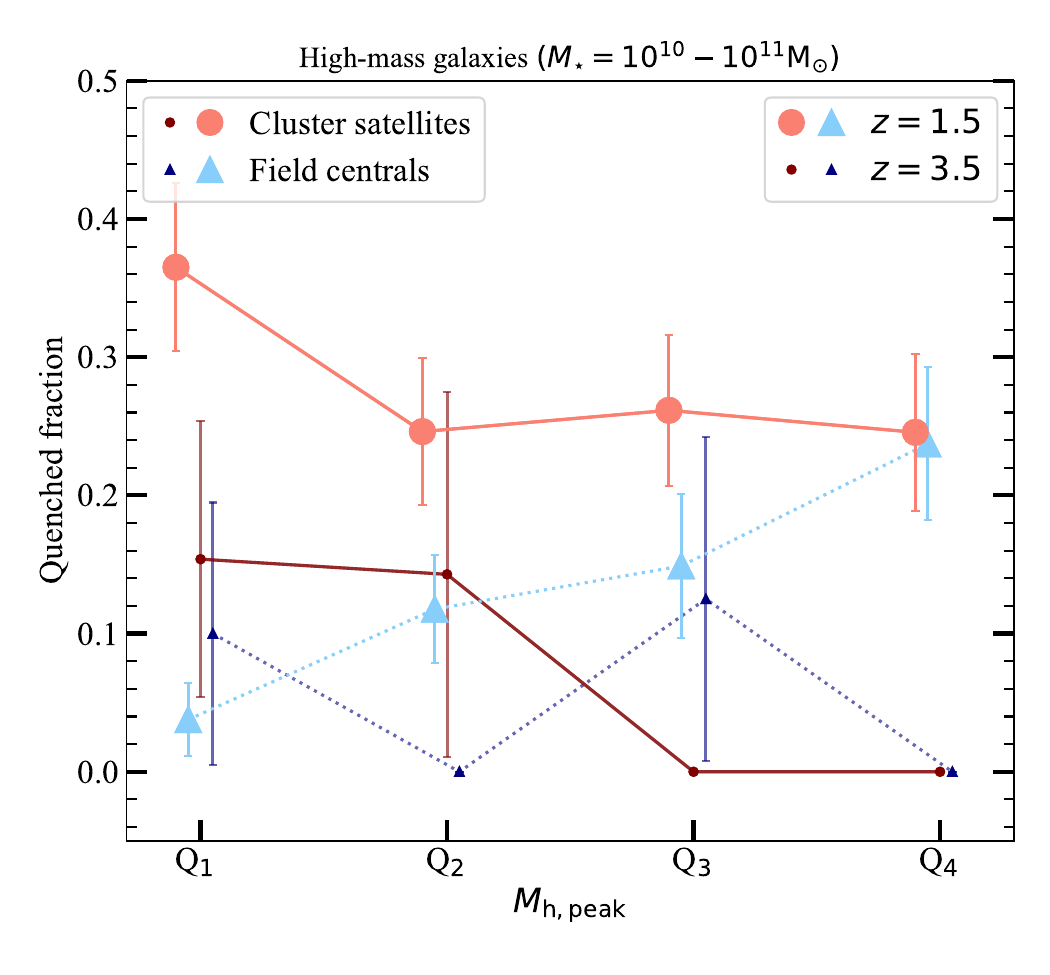}
        \caption{Fraction of quenched galaxies vs.~the $M_{\mathrm{h, peak}}$ quartile in clusters (red shades) and field (blue shades) at different redshifts. The evolution is shown for high-mass galaxies, with $10^{10} - 10^{11} \msun$. Higher redshift data points are respectively smaller and in darker tones of the same colours. Similar to the left panel of Fig.~\ref{fig:qf_vs_vmax_range}, at $z=3.5$, field centrals and cluster satellites have a comparable trend of quenched galaxy fraction. Also, for the high-mass field centrals, a higher $M_{\mathrm{h, peak}}$ corresponds to a higher quenched fraction at $z=1.5$, as observed in Fig.~\ref{fig:qf_vs_vmax_range}. At the highest quartile, the field centrals and cluster satellites have comparable quenched fraction as well. However, the enhancement of quenched fraction is stronger at lower halo-mass quartiles, compared to lower $v_{\mathrm{max, peak}}$ quartiles.}
        \label{fig:qf_mh}
    \end{figure}
\end{center}

Instead of the $\vmaxpeak$, here we checked the quenched fraction of cluster satellites and field centrals in $M_{\mathrm{h, peak}}$ quartiles. Figure~\ref{fig:qf_mh} shows this for the high-mass ($\mstar$ within $10^{10} - 10^{11} \msun$) galaxy sample in clusters and the field. This is analogous to the left panel of Fig.~\ref{fig:qf_vs_vmax_range}, excluding the data points at $z=2$ to demonstrate the most trend between our highest and lowest redshifts of concern. 

Similar to the left panel of Fig.~\ref{fig:qf_vs_vmax_range}, at $z=3.5$, cluster satellites and field centrals have a comparable quenched fraction against the $M_{\mathrm{h, peak}}$ quartiles, which again supports the notion that at this redshift, cluster and field galaxies likely quenched through similar mechanisms. At $z=1.5$, field centrals have a similar trend as in Fig.~\ref{fig:qf_vs_vmax_range} -- having a higher quenched fraction in higher halo mass quartiles. Cluster satellites, however, differ strongly at the lowest halo mass quartile. At the other three points, the trend is comparable to Fig.~\ref{fig:qf_vs_vmax_range}, especially at the highest quartile, where their quenched fraction is the same as the field centrals. Overall, Fig.~\ref{fig:qf_mh} is consistent with our main conclusions about the high-mass galaxies: at this stellar mass range, if a galaxy has a higher halo mass ($Q_4$), it is similarly likely to be quenched in both clusters and the field, but for lower halo masses, it is more likely to be quenched as a satellite in a massive cluster than an isolated galaxy in the field.

\section{Black hole mass, half mass radius, and cluster-centric distance}
\label{sec:app_mbh_r_half}

\begin{center}
    \begin{figure}
        \includegraphics[width= \columnwidth]{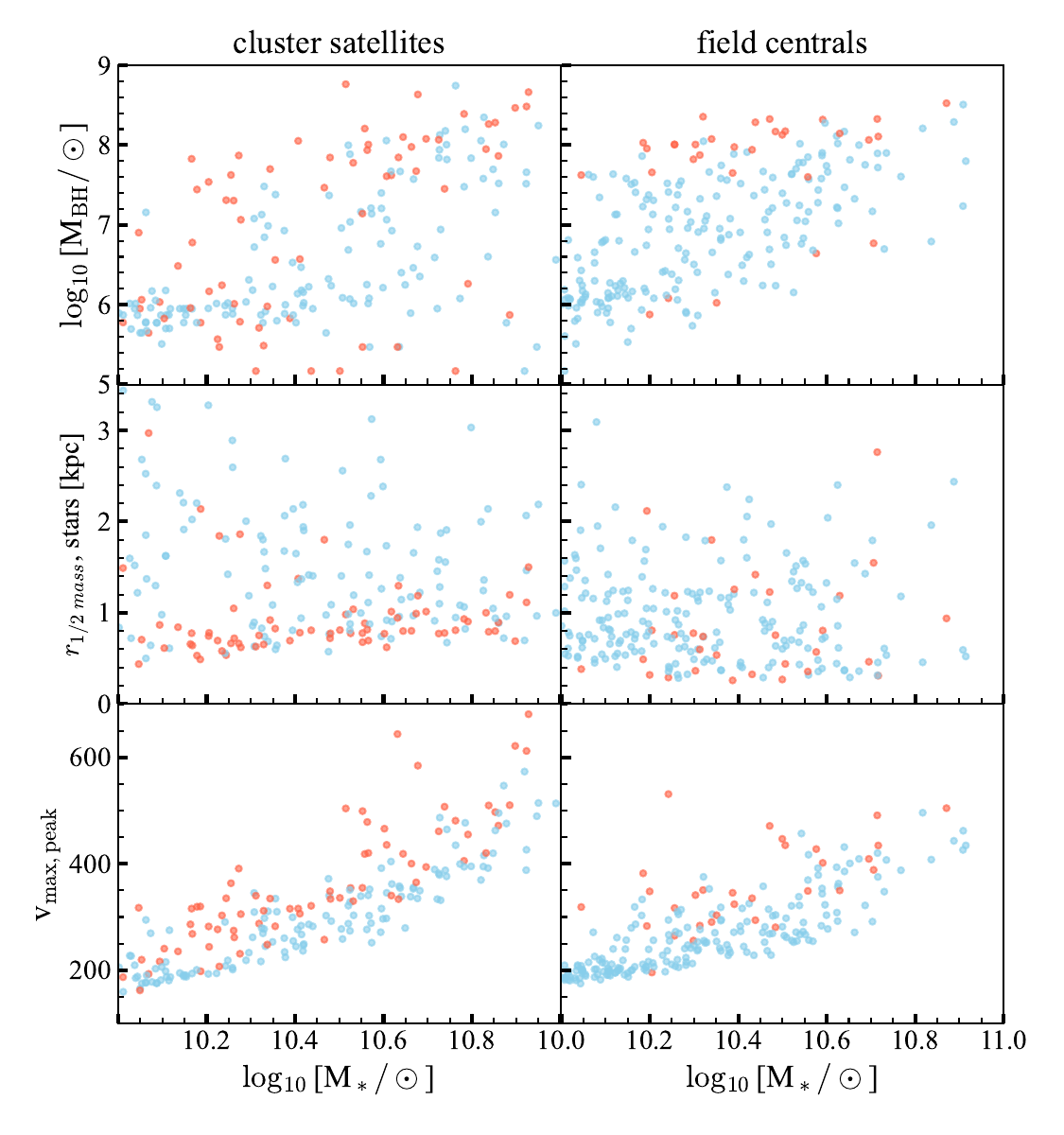}
        \caption{Central SMBH mass (top), half mass radius (middle), and $v_{\mathrm{max, peak}}$ (bottom) vs the stellar mass of cluster satellites (left panels) and field centrals (right panels) in galaxy stellar mass range $10^{10} - 10^{11} \msun$. Quenched (sSFR$\leq$-10) and star-forming galaxies are in red and blue, respectively. Quenched cluster satellites have a higher $v_{\mathrm{max, peak}}$, smaller ($<1$~kpc) half-mass radii, and a high scatter in SMBH mass range compared to the star-forming ones. Quenched field centrals also have a higher $v_{\mathrm{max, peak}}$ value, however, unlike the cluster satellites, their half-mass radii have a higher scatter whereas the SMBH masses have a consistent higher value ($\geq 10^{10}~\msun$).}
        \label{fig:mbh_r_half_vmax}
    \end{figure}
\end{center}

We studied the distribution of $M_{\mathrm{SMBH}}$ and $r_{\mathrm{1/2, mass}}$ with $M_*$ of our high-mass galaxy samples for field centrals and cluster satellites and compared these trends with those as a function of $v_{\mathrm{max, peak}}$ as used in the main text. This is shown in Fig.~\ref{fig:mbh_r_half_vmax}. For all the panels here, star-forming (blue) and quenched (red) galaxies have the same selection criteria as is throughout the paper. For both cluster satellites and field centrals, most of the quenched galaxies have a high $v_{\mathrm{max, peak}}$, which we discussed in detail in Sec.~\ref{sec:result}. However, we see different distributions of quenched cluster and field galaxies in terms of the $M_{\mathrm{SMBH}}$ and $r_{\mathrm{1/2, mass}}$. 
Quenched cluster satellites have a smaller $r_{\mathrm{1/2, mass}}$ ($\leq$1kpc) for 74 per cent of the sample galaxies, and 90 per cent of the quenched cluster satellites have their $r_{\mathrm{1/2, mass}}$  value $\leq$ 1.5kpc. In terms of their $M_{\mathrm{SMBH}}$, there is a larger scatter -- 41 per cent of them have a high SMBH mass ($\geq 10^{7.5}\msun$), but the rest have a wide range of values, with some as low as $10^5 \msun$. On the other hand, 85 per cent of the quenched field centrals have a high $M_{\mathrm{SMBH}}$ value ($\geq 10^{7.5} \msun$), with a few having mass $\sim 10^6 \msun$. In terms of their $r_{\mathrm{1/2, mass}}$, they have a large scatter with 67 per cent having $r_{\mathrm{1/2, mass}}\leq$1~kpc. The high fraction of quenched field centrals with a high SMBH mass is consistent with the scenario that they are self-quenched and the process is primarily AGN feedback, which can get stronger for more massive black holes. 

With cluster satellites, however, both the smaller size and more massive SMBH of the cluster satellites could be connected to their previous proximity to the cluster halo centre. In that case, their quenching that followed their proximity to the cluster centre has a high chance of being environment-driven. To further test the effect of environment on cluster satellites, we measured the distance of these satellites from the cluster centre between $z = 14$ and the epoch when $\vmaxpeak$ occurs, in units of $r_{200}$. If a satellite came too close to the central galaxy before $\vmaxpeak$ occurred, and therefore, was subjected to hydrodynamical and/or tidal forces, then it could possibly be stripped of its cold gas, and consequently quench its star formation. 

Instead of the standard 500~Myr time-steps of the simulation outputs, we used a smaller 10~Myr time-step to reduce the chance of missing a short-lived phase where a satellite may be the closest to the central galaxy. We plotted the minimum distance between the satellite and central over the considered redshift range vs the stellar mass of the galaxies (similar to Fig.~\ref{fig:mbh_r_half_vmax} but not shown here). We found that only 27 per cent of the quenched satellites have come as close as $0.2\times r_{200}$ of the cluster at some point before attaining their $\vmaxpeak$. Therefore, stripping in the cluster environment cannot be the primary reason for quenching the cluster satellites. 

Moreover, half of the quenched satellites were never within $r_{200}$ distance from the cluster centres before $\vmaxpeak$ occurred. These satellites were likely already self-quenched before being a part of the cluster. A similar suggestion was made by \citet{werner2022}, where they considered the area within $1<r/r_{200}<3$ distances of GOGREEN clusters as the cluster infall region. Because of the excess quenching in the infall region, they suggested that some massive quenched galaxies in the infall region quenched before they became part of the clusters. In Hydrangea, galaxies in $1<r/r_{200}<3$ distances can also be part of the cluster halo if they satisfy the FoF membership criteria. In our sample, most of the massive quenched cluster satellites with $M_{\mathrm{SMBH}}\geq 10^{7.5}\,\msun$, have a minimum distance from the cluster centre (before attaining $\vmaxpeak$) above their corresponding $r_{200}$. They are, therefore, in the `infall region' defined by \citet{werner2022}. This can also indicate that these quenched cluster galaxies with a massive black hole are central galaxies of infalling groups. \citet{baxter2023} also found that pre-processing plays an important role in reproducing the observed quenched fraction of massive satellites in massive $z>1$ clusters. Another of their finding, that ram-pressure stripping only contributes as a secondary process, is also consistent with our findings, given that only a small fraction of the quenched satellites in our sample come close enough to the cluster centres to undergo ram-pressure striping. However, the likely dominant quenching mechanism from their findings, starvation, was not tested in our sample as it is out of the scope of this work. %Therefore, some of them have a high-mass SMBH but most of them have a smaller $r_{\mathrm{1/2, mass}}$. 

In total, 97 per cent of the quenched field centrals have either $r_{1/2, \mathrm{mass}}\leq$1 kpc or $M_{\mathrm{SMBH}}\geq 10^{7.5}\msun$, whereas, 87 per cent of the quenched cluster satellites satisfy the same conditions or have been at a distance of less than 20\% of the $r_{200}$ from the cluster centre at some point before $\vmaxpeak$ occurred. Therefore, the majority of quenched galaxies in our samples are consistent with having a high $v_{\mathrm{max, peak}}$, which is connected to the halo mass, with a combination to the SMBH mass, galaxy size, and proximity to the cluster centre (only for the satellites). Dynamical interactions can be another important factor for the quenched galaxies, especially for the remaining 13 per cent cluster satellites and 3 per cent field centrals that do not correlate with any of our tested parameters. 

%%%%%%%%%%%%%%%%%%%%%%%%%%%%%%%%%%%%%%%%%%%%%%%%%%

% Don't change these lines
\bsp	% typesetting comment
\label{lastpage}
\end{document}